\documentclass[a4paper,fleqn,usenatbib]{mnras}

\usepackage[T1]{fontenc}
\usepackage{ae,aecompl}

\usepackage{graphicx}   
\usepackage{amsmath}    
\usepackage{amssymb}    

\newcommand{\hii}{\mbox{H\,\textsc{ii}}} 
\newcommand{\hi}{\mbox{H\,\textsc{i}}} 

\title[A new SNR G21.8$-$3.0]{Discovery of a new supernova remnant
  G21.8$-$3.0}

\author[X. Y. Gao et al.]{
X. Y. Gao,$^{1,2}$\thanks{E-mail: xygao@nao.cas.cn (XYG)}
P. Reich,$^{3}$
W. Reich,$^{3}$
L. G. Hou$^{1,2}$
and J. L. Han,$^{1,2,4}$
\\
$^{1}$National Astronomical Observatories, CAS, Jia-20 Datun Road,
Chaoyang District, Beijing 100101, PR China \\
$^{2}$CAS Key Laboratory of FAST, National Astronomical Observatories,
Chinese Academy of Sciences \\
$^{3}$Max-Planck-Institut f\"{u}r Radioastronomie, Auf dem H\"{u}gel
69, 53121 Bonn, Germany \\
$^{4}$School of Astronomy, University of Chinese Academy of Sciences,
Beijing 100049, China
}

\date{Accepted XXX. Received YYY; in original form ZZZ}

\pubyear{2020}

\begin{document}
\label{firstpage}
\pagerange{\pageref{firstpage}--\pageref{lastpage}}
\maketitle

\begin{abstract}
Sensitive radio continuum surveys of the Galactic plane are ideal for
discovering new supernova remnants (SNRs). From the Sino-German
$\lambda$6\ cm polarisation survey of the Galactic plane, an extended
shell-like structure has been found at $\ell = 21\fdg8, b = -3\fdg0$,
which has a size of about 1$\degr$. New observations were made with
the Effelsberg 100-m radio telescope at $\lambda$11\ cm to estimate
the source spectrum together with the Urumqi $\lambda$6\ cm and the
Effelsberg $\lambda$21\ cm data. The spectral index of G21.8$-$3.0 was
found to be $\alpha = -0.72\pm0.16$. Polarised emission was mostly
detected in the eastern half of G21.8$-$3.0 at both $\lambda$6\ cm and
$\lambda$11\ cm. These properties, together with the H$\alpha$
filament along its northern periphery and the lack of infrared
emission, indicate that the emission is non-thermal as is usual in
shell-type SNRs.
\end{abstract}




\section{Introduction}

Supernova explosions process the Galactic interstellar medium (ISM)
and enhance dynamics in the Galaxy by e.g. accelerating particles,
enriching chemical elements, and triggering the next generation of
star formation \citep{Padmanabhan01}. The remnants glow for more than
hundreds of thousands years when interacting with the ISM. The total
count of these Galactic supernova remnants (SNRs), as a basic
parameter, is however still unclear. A large deficiency exists between
the predicted \citep{Li91, Tammann94} and the observed number
\citep{Ferrand12, Green19}. To bridge this gap, extensive discoveries
of new SNRs depend on deeper radio continuum surveys to unveil the
faint cases, and sharper views to separate small structures,
especially toward the inner Galaxy \citep[e.g.][]{Brogan06}. Recently,
a big step was made by \citet{Anderson17}: through combining the radio
continuum data from ``The $\hi$, OH, Recombination line survey of the
Milky Way'' (THOR, angular resolution $\sim$20$\arcsec$), the VLA 1.4
GHz Galactic Plane Survey \citep[VGPS, angular resolution
  $\sim$1$\arcmin$,][]{Stil06}, and the Effelsberg $\lambda$21\ cm
survey \citep[][angular resolution $\sim$9$\arcmin$]{Reich9021}, they
identified 76 new SNR candidates in the area of $17\fdg5 < \ell <
67\fdg4$ and $|b| \leqslant 1\fdg25$. A number of new SNRs have also
been identified through very careful inspections in arcmin-resolution
images of very sensitive large-scale radio surveys
\citep[e.g.][]{Foster13, Kothes14, Gerbrandt14, Kothes17} and
occasionally from deep H$\alpha$ observations \citep[e.g.][]{Fesen10,
  Fesen15}.

Using the data of the Sino-German $\lambda$6\ cm polarisation survey
of the Galactic plane\footnote{Data available at
  http://www.mpifr-bonn.mpg.de/survey.html and
  http://zmtt.bao.ac.cn/6cm/surveydata.html} \citep{Sun07, Gao10,
  Sun11a, Xiao11}, we discovered three new SNRs: G25.1$-$2.3 and
G178.2$-$4.2 by \citet{Gao11y} and G150.3+4.5 by \citet{Gao14}. Here
we report the discovery of a new shell-like SNR, G21.8$-$3.0. We
introduce the basic data in Sect.~2, and discuss the properties of
G21.8$-$3.0 in Sect.~3. A summary is given in Sect.~4.

\section{Basic data}

\subsection{Urumqi $\lambda$6\ cm data}

The $\lambda$6\ cm total-intensity (Stokes $I$) and
linear-polarisation (Stokes $Q$ and $U$) data were extracted from the
Sino-German $\lambda$6\ cm polarisation survey of the Galactic
plane. The data were acquired by the Urumqi 25-m radio telescope of
Xinjiang Astronomical Observatory, equipped with a single-channel
receiver constructed by the Max-Planck-Institut f{\"u}r
Radioastronomie, Germany. The angular resolution of the survey is
9$\farcm$5. The sensitivity is about 1.0~mK $T_{\rm B}$ for $I$ and
0.5~mK $T_{\rm B}$ for $Q$/$U$ in the inner Galactic part of the
survey. Details on the $\lambda$6\ cm receiving system, observation
strategy, set-up and data reduction were described in \citet{Sun06,
  Sun07} and \citet{Gao10}.  SNRs with surface brightness larger than
$\rm \Sigma_{1~GHz} \sim 3.9 \times 10^{-23}$ [${\rm
    Wm^{-2}Hz^{-1}sr^{-1}}$] (3$\sigma$ detection limit) could be
detected in this survey \citep{Gao10}.

An extended shell-like source, G21.8$-$3.0 (see Fig.~\ref{G21.8}),
with associated polarised emission was found at $\ell = 21\fdg8, b =
-3\fdg0$, which has a size of about 1$\degr$.

\subsection{Effelsberg $\lambda$11\ cm data}

To reveal the properties of G21.8$-$3.0, new $\lambda$11\ cm
total-intensity and linear-polarisation observations were made with
the Effelsberg 100-m radio telescope in October, 2018. We decided not
to use the standard 8-channel 80-MHz bandwidth backend centred at
2639.5 MHz, with which tests in December 2017 showed that strong radio
interference (RFI) reduced the usable bandwidth to 40~MHz. Instead,
the 2-GHz wide SPECPOL polarimeter with 1024 channels available at the
Effelsberg telescope was successfully attached to the receiver to
check the RFI situation outside the standard 80-MHz band. Although the
spectrometer covered the band from 2300 MHz to 4300 MHz, the filter in
the $\lambda$11\ cm frontend limits the accessible frequency range
from 2599.5~MHz to 2679.5~MHz. The test was successful and showed that
in total four frequency bands were usable. The remaining RFI could be
removed either by software or by data editing. The four bands cover in
total about 185~MHz, which improves on the bandwidth available so
far. The four bands were centred at 2376.25~MHz, 2553.75~MHz,
2658.0~MHz, and 2696.25~MHz and have a usable bandwidth of 32.5~MHz,
117.5~MHz, 8~MHz, and 27.5~MHz, respectively. The NOD2-based data
reduction of G21.8$-$3.0 followed the same routines as for the
$\lambda$6\ cm data \citep{Gao10}. The angular resolutions of the four
sections are slightly different, from 4$\farcm$2 to 4$\farcm$7. After
data reduction, we added them after convolving all the images to
4$\farcm$8. The sensitivity of the combined data is about 5.0~mK
$T_{\rm B}$ and 4.5~mK $T_{\rm B}$ for $I$ and $Q$/$U$,
respectively. We note, however, that the final noise does not fully
reflect the increase in bandwidth when compared to earlier
$\lambda$11\ cm observations. This indicates either the existence of
low-level RFI in- or out-side the selected bands and requires further
tests for optimisation in the future.

\subsection{Effelsberg $\lambda$21\ cm data}

The $\lambda$21\ cm total-intensity data of G21.8$-$3.0 came from the
Effelsberg survey \citep{Reich9021}. Polarisation data were not
available. The central observing frequency of the survey was
1408~MHz. The bandwidth was often set to 20~MHz. The beam size is
9$\farcm$4, nearly identical to that of the Urumqi $\lambda$6\ cm
survey. Typical r.m.s. of the noise was about 40~mK $T_{\rm B}$. The
high angular resolution Canadian Galactic Plane Survey
\citep[CGPS,][]{Landecker10} and VGPS \citep{Stil06}, which surveyed
the Galactic plane both in radio continuum and $\hi$, do not cover this
area.

\subsection{Other data}

Besides the radio continuum data, additional data were also used to
study the properties of G21.8$-$3.0. The optical H$\alpha$ data were
taken from the Southern H$\alpha$ Sky Survey Atlas \citep[hereafter
  SHASSA,][]{Gaustad01}. Two sets of infrared data around G21.8$-$3.0
were adopted for comparison with the radio continuum emission. The
60$\mu$m data were from the Improved Reprocessing of the Infrared
Astronomical Satellite Survey \citep[IRIS,][]{Miville05}. The 12$\mu$m
and 22$\mu$m WISE data \citep{Wright10} and the WISE $\hii$ region
catalogue\footnote{http://astro.phys.wvu.edu/wise/} \citep{Anderson14}
were also consulted. In addition, the X-ray data from ROSAT all-sky
survey \citep{Voges99} and Chandra
Observatory\footnote{https://cda.harvard.edu/chaser/}, $\gamma$-ray
sources identified in the HESS and HAWC \citep{Jardin19} have also
been checked for a high-energy counterpart associated with
G21.8$-$3.0. $\hi$ data from the Parkes Galactic all-sky survey \citep
{GASS1,GASS2} and the $^{12}$CO($J=1-0$) data from \citet{dht01} were
checked for a possible coincident structure to estimate the distance
of G21.8$-$3.0.

\begin{figure*}
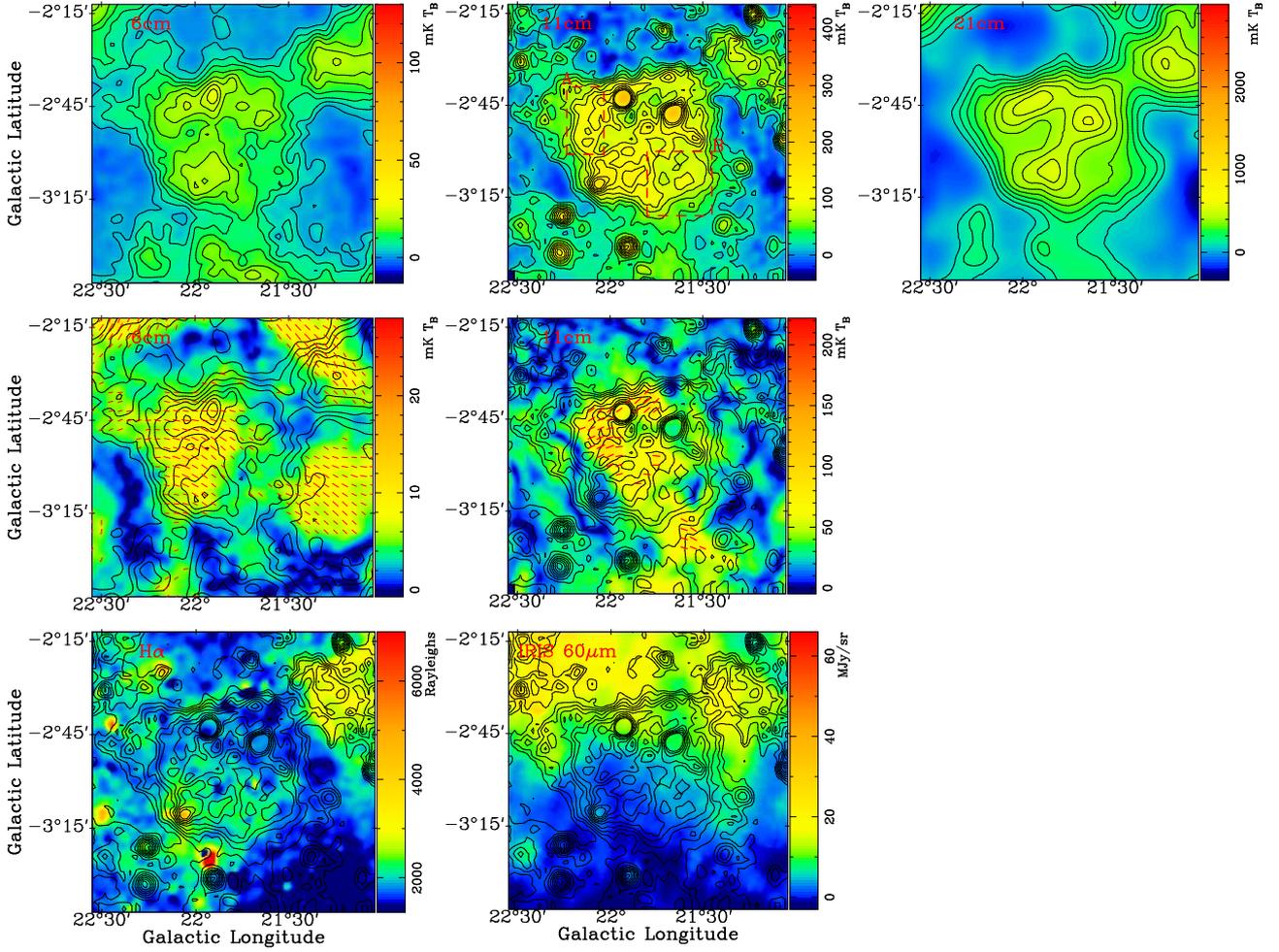

\begin{tabular}{lll}
  \includegraphics[angle=-90, width=0.33\textwidth]{G21.8-3_6cm1.ps}
  \includegraphics[angle=-90, width=0.31\textwidth]{G21.8-3_11cm1.ps}
  \includegraphics[angle=-90, width=0.31\textwidth]{G21.8-3_21cm1.ps}\\
  \includegraphics[angle=-90, width=0.33\textwidth]{G21.8-3_6pi1.ps}
  \includegraphics[angle=-90, width=0.31\textwidth]{G21.8-3_11pi1.ps}\\
  \includegraphics[angle=-90, width=0.33\textwidth]{G21.8-3_Ha1.ps}
  \includegraphics[angle=-90, width=0.31\textwidth]{G21.8-3_infra1.ps}\\
\end{tabular}
\caption{Radio continuum images of G21.8$-$3.0. {\it Top panels from
    left to right}: total intensity $I$ of G21.8$-$3.0 from the Urumqi
  $\lambda$6\ cm, Effelsberg $\lambda$11\ cm, and $\lambda$21\ cm
  observations. The $\lambda$6\ cm and $\lambda$21\ cm data were
  extracted from the legacy Galactic plane surveys, while the
  $\lambda$11\ cm data were newly obtained as described in
  Sect.~2.2. The contours run at 3.0 + (n-1) $\times$ 3.0~mK $T_{\rm
    B}$, at 15.0 + (n-1) $\times$ 15.0~mK\ $T_{\rm B}$, and 90 + (n-1)
  $\times$ 60~mK\ $T_{\rm B}$ (n = 1,2,3 ...) in the $\lambda$6\ cm,
  $\lambda$11\ cm, and $\lambda$21\ cm images, respectively. The two
  dashed rectangles in the $\lambda$11\ cm image mark the regions for
  TT-plots. {\it Middle panels from left to right:} polarised
  intensity $PI$ of G21.8$-$3.0 at $\lambda$6\ cm and
  $\lambda$11\ cm. The contours are the same as in the {\it top
    panels}. {\it Bottom panels from left to right:} SHASSA H$\alpha$
  and IRIS 60$\mu$m images of G21.8$-$3.0 overlaid by $\lambda$11\ cm
  contours.}
\label{G21.8}
\end{figure*}

\begin{table}
\renewcommand\tabcolsep{13.5pt}
\begin{center}
\caption{Flux densities of three strong point-like sources overlapping
  G21.8$-$3.0.}
\label{ps}
\begin{tabular}{cccc}
\hline\hline
\multicolumn{1}{c}{Source}  &\multicolumn{1}{c}{$S_{\rm \lambda21\ cm}$}  &\multicolumn{1}{c}{$S_{\rm \lambda11\ cm}$}   &\multicolumn{1}{c}{$S_{\rm \lambda6\ cm}$} \\
\multicolumn{1}{c}{}  &\multicolumn{1}{c}{(mJy)}  &\multicolumn{1}{c}{(mJy)}   &\multicolumn{1}{c}{(mJy)}  \\
\hline
G22.08$-$3.19    & 37.7   &  35.5$^{*}$  & 33.3$^{*}$  \\
G21.93$-$2.72    & 115.6  &  77.7  & 64   \\
G21.66$-$2.79    & 102.0  &  83.6  & 67.9$^{*}$  \\
\hline
%
\end{tabular}
\end{center}
{Notes: values with $^{*}$ are extrapolated.}
\end{table}

\section{Properties of G21.8$-$3.0}

\subsection{Radio structure, polarisation and spectral index}

\begin{figure*}
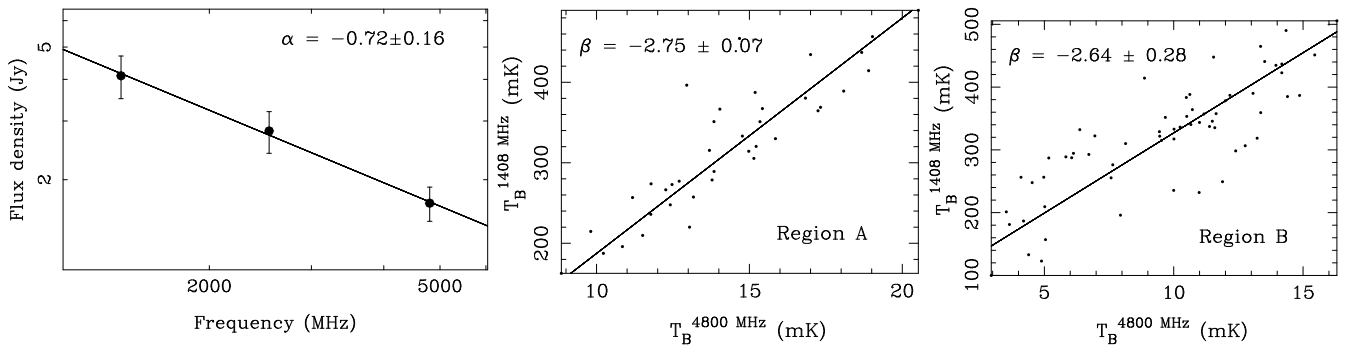

\begin{tabular}{ccc}
  \includegraphics[angle=-90, width=0.358\textwidth]{spec.ps}
  \includegraphics[angle=-90, width=0.315\textwidth]{east.ps}
  \includegraphics[angle=-90, width=0.305\textwidth]{west.ps}
\end{tabular}
  \caption{{\it Left panel}: radio continuum spectrum of G21.8$-$3.0
    fitted by the flux densities obtained from the Urumqi
    $\lambda$6\ cm data, the Effelsberg $\lambda$11\ cm and
    $\lambda$21\ cm data after removing the contributions from three
    point-like sources listed in Table~\ref{ps}. {\it Middle and right
      panels:} TT-plots by using the Urumqi $\lambda$6\ cm data
    against Effelsberg $\lambda$21\ cm data for the Region A ({\it
      middle panel}) and B ({\it right panel}) as indicated in the
    {\it top middle panel} of Fig.~\ref{G21.8}.}
\label{TT}
\end{figure*}

We show total-intensity ($I$) images of G21.8$-$3.0 from the Urumqi
$\lambda$6\ cm survey, the new Effelsberg $\lambda$11\ cm observations
(a combination of the four individual bands) and the archived
Effelsberg $\lambda$21\ cm survey in the {\it top panels} of
Fig.~\ref{G21.8}.  With the higher angular resolution of 4$\farcm$8,
more details of G21.8$-$3.0 are visible at
$\lambda$11\ cm. G21.8$-$3.0 has a nearly circular shape with a
diameter of about 1$\degr$. Enhanced total-intensity emission is seen
in a partial shell structure extending from northeast to north and a
small circular section in the southeast. Three point-like sources were
identified: G21.93$-$2.72 with a spectral index of $\alpha = -0.63$
\citep{Vollmer10}, the planetary nebula G22.08$-$3.19 \citep{Frew13},
and G21.66$-$2.79 about which not much information is available in
literature. We list their flux densities at $\lambda$21\ cm,
$\lambda$11\ cm, and $\lambda$6\ cm in Table~\ref{ps}. All $S_{\rm
  \lambda21\ cm}$ values are retrieved from the NVSS source catalogue
\citep{Condon98}. The $S_{\rm \lambda11\ cm}$ values of G21.93$-$2.72
and G21.66$-$2.79 are derived from Gaussian fits to the newly observed
Effelsberg $\lambda$11\ cm data and the $S_{\rm \lambda6\ cm}$ value
for G21.93$-$2.72 is taken from the PWN survey \citep{Wright94}. Their
contributions were subtracted so that the flux densities of the
extended source G21.8$-$3.0 at these three bands can be estimated.

The $PI$ (Polarised Intensity, $PI = \sqrt{Q^{2} + U^{2}}$) images at
$\lambda$6\ cm and $\lambda$11\ cm are shown in the {\it middle
  panels} of Fig.~\ref{G21.8}. The orientation of $B$-field vectors of
the polarised emission seems to be uniform, implying the presence of
regular magnetic fields. Strong polarised emission is seen outside
G21.8$-$3.0 at $\lambda$6\ cm without any corresponding
total-intensity counterparts. These patches are either Faraday screens
or caused by the turbulent magnetic fields as discussed in
\citet{Sun11a}. By comparing the polarisation angles ($PA$s, $PA =
\frac{1}{2}atan{\frac{U}{Q}}$) at these two bands, the rotation
measure ($RM$, $RM = \Delta PA / ({\lambda_{\rm
    6cm}}^{2}-{\lambda_{\rm 11cm}}^{2})$) of the source can be
estimated. For the region with prominent polarised emission, i.e. the
eastern part of G21.8$-$3.0, we obtained an averaged $RM$ of
$\sim$80~rad\ m$^{-2}$ with n$\pi$ ambiguity of 312~rad\ m$^{-2}$ from
the $\lambda$6\ cm data and the combined $\lambda$11\ cm data (the
central frequency was taken as 2536.25~MHz) convolved to
9$\farcm$5. This result was tested by the data at the high end
(2696.25~MHz) and the low end (2376.25~MHz) of the new $\lambda$11\ cm
observations at the angular resolution of 4$\farcm$8. We got a
consistent $RM$ of $\sim$90~rad\ m$^{-2}$. The n$\pi$ ambiguity is
about $\pm$864~rad\ m$^{-2}$, which seems very unlikely and implies
that $RM$ $\sim$ 80 -- 90\ rad\ m$^{-2}$ of G21.8$-$3.0 is most likely
convincing. No $RM$s of extra-Galactic sources behind G21.8$-$3.0 are
available to diminish the ambiguity \citep{Xu14}.

The radio continuum spectral index is a key to distinguish the
non-thermal synchrotron emission from SNRs and thermal bremsstrahlung
radiation from $\hii$ regions. Most shell-type SNRs have a spectral
index around $\alpha \sim -0.5$ ($S_{\nu} \sim \nu^{\alpha}$)
\citep[see][]{Urosevic14, Dubner17}, which makes them distinct from
the optically-thin emission of $\hii$ regions with a much flatter
spectrum of $\alpha \sim -0.1$ \citep[see][]{Wilson13}. To estimate
the flux densities of G21.8$-$3.0 at each frequency, a twisted plane
was fitted and subtracted in each map to remove the unrelated
foreground/background emission before integration. We got $S_{\rm
  \lambda6\ cm}$ = 1.7$\pm$0.2~Jy, $S_{\rm \lambda11\ cm}$ =
2.8$\pm$0.4~Jy, and $S_{\rm \lambda21\ cm}$ = 4.1$\pm$0.6~Jy for
G21.8$-$3.0 after the point-like sources were subtracted.  The radio
continuum spectrum was fitted and is shown in the {\it left panel} of
Fig.~\ref{TT}. The spectral index is $\alpha = -0.72\pm0.16$, which
clearly indicates the non-thermal nature of G21.8$-$3.0.

The TT (Temperature - Temperature) plot method introduced by
\citet{Turtle62} has been widely used to reveal the spectral index of
an extended emission because it is insensitive to zero-level
uncertainties of images. Taking the notion that $T_{{\rm source}}(\nu)
\sim \nu^\beta$, where $T_{\rm source}$ is the temperature of the
target source and $\beta$ is defined as the brightness temperature
spectral index, the temperature map observed at frequency $\nu$ can be
described as $T(\nu) = T_{\rm source}(\nu) + C$. Here, $C$ is the
baselevel offset. By substituting $T_{\rm source}$, it is then
straightforward to have $T(\nu_{1}) =
(\frac{\nu_{1}}{\nu_{2}})^{\beta}\ T(\nu_{2}) -
(\frac{\nu_{1}}{\nu_{2}})^{\beta}\ C_{2} + C_{1}$, implying that the
slope of linear regression in TT-plots can be used to derive source
spectral index $\beta$ and the baselevel offsets included in the
intercept do not have any influence on the slope. Considering the flux
density $S \sim 2kT\nu^{2} \sim \nu^{\alpha}$ for the radio image and
$T \sim \nu^{\beta}$, one can relate the flux density and brightness
temperature spectral indices as $\alpha = \beta + 2$. For G21.8$-$3.0,
two areas were selected (marked as A and B in the {\it top middle
  panel} of Fig.~\ref{G21.8}) for TT-plots, which are free from
point-like source contamination by inspecting the high angular
resolution NVSS image. Region A is the shell-like structure in the
northeastern part of G21.8$-$3.0, and Region B is an area which is
relatively weak in emission. The TT-plots give the brightness
temperature spectral indices as being $\beta = -2.75\pm0.07$ and
$\beta = -2.64\pm0.28$ between the $\lambda$6\ cm and $\lambda$21\ cm
data for Region A and B, respectively, very consistent with the result
derived from the flux densities at the three bands.

Based on the flux density measured at $\lambda$6\ cm and the fitted
spectral index of $\alpha = -0.72$, we estimated the surface
brightness of G21.8$-$3.0 at 1~GHz to be $\rm \Sigma_{1~GHz} =
2.2\pm0.3 \times 10^{-22}$ [${\rm Wm^{-2}Hz^{-1}sr^{-1}}$], placing it
at the fainter end of the known Galactic SNRs \citep[see Fig.~1
  in][]{Green14}.

\subsection{Properties at optical, infrared and high-energy bands}

Some SNRs do show features at optical and infrared bands, and a few
were even newly identified from optical observations, e.g. SNRs
G159.6+7.3 and G70.0$-$21.5 discovered through deep H$\alpha$
observations \citep{Fesen10, Fesen15}.

We extracted the continuum-corrected H$\alpha$ emission from SHASSA
\citep{Gaustad01} for the G21.8$-$3.0 field as shown in the {\it
  bottom left panel} of Fig.~\ref{G21.8}. The image was slightly
smoothed to 4$\arcmin$ from the original 3$\farcm$2 resolution. For
the planetary nebula G22.08$-$3.19, bright H$\alpha$ emission is
visible. Some diffuse H$\alpha$ emission overlaps the lower part of
G21.8$-$3.0, however, no solid correlation can be established. The
most interesting feature seen in the H$\alpha$ image is an elongated
filament nicely running along the northern edge of G21.8$-$3.0. It is
necessary to confirm with other optical lines, such as
$[\mbox{S\,\textsc{ii}}]$ and $[\mbox{O\,\textsc{i}}]$
\citep[e.g.][]{Fesen85, Kopsacheili19} whether it is a shock emission
filament.

The IRIS 60$\mu$m image of G21.8$-$3.0 is shown in the {\it bottom
  middle panel} of Fig.~\ref{G21.8}. The WISE images, which we did not
present, resemble the structures in the IRIS map, but show more point
sources due to their higher angular resolution. No catalogued WISE
$\hii$ regions \citep{Anderson14} are found in this region. SNRs have
faint far-infrared emission compared to $\hii$ regions \citep{Fuerst87a}
and lack mid-infrared emission \citep{Anderson17}.  No clear
morphological correlation is found between the infrared (IRIS and
WISE) and radio images of G21.8$-$3.0. Based on the comparison between
the 60$\mu$m infrared and Effelsberg $\lambda$11\ cm data,
\citet{Fuerst87a} found that the infrared to radio continuum ratio $R$
is as high as 1000$\pm$500 for $\hii$ regions, and proposed that the
$\lambda$11\ cm sources with $R$ < 250 could be SNR candidates. We
found low $R$ values, i.e.  $R \sim$ 80 in the lower part and $R \sim$
240 in the upper part of G21.8$-$3.0, which support its SNR nature.

\begin{figure*}
  \centering
  \includegraphics[width=0.8\textwidth]{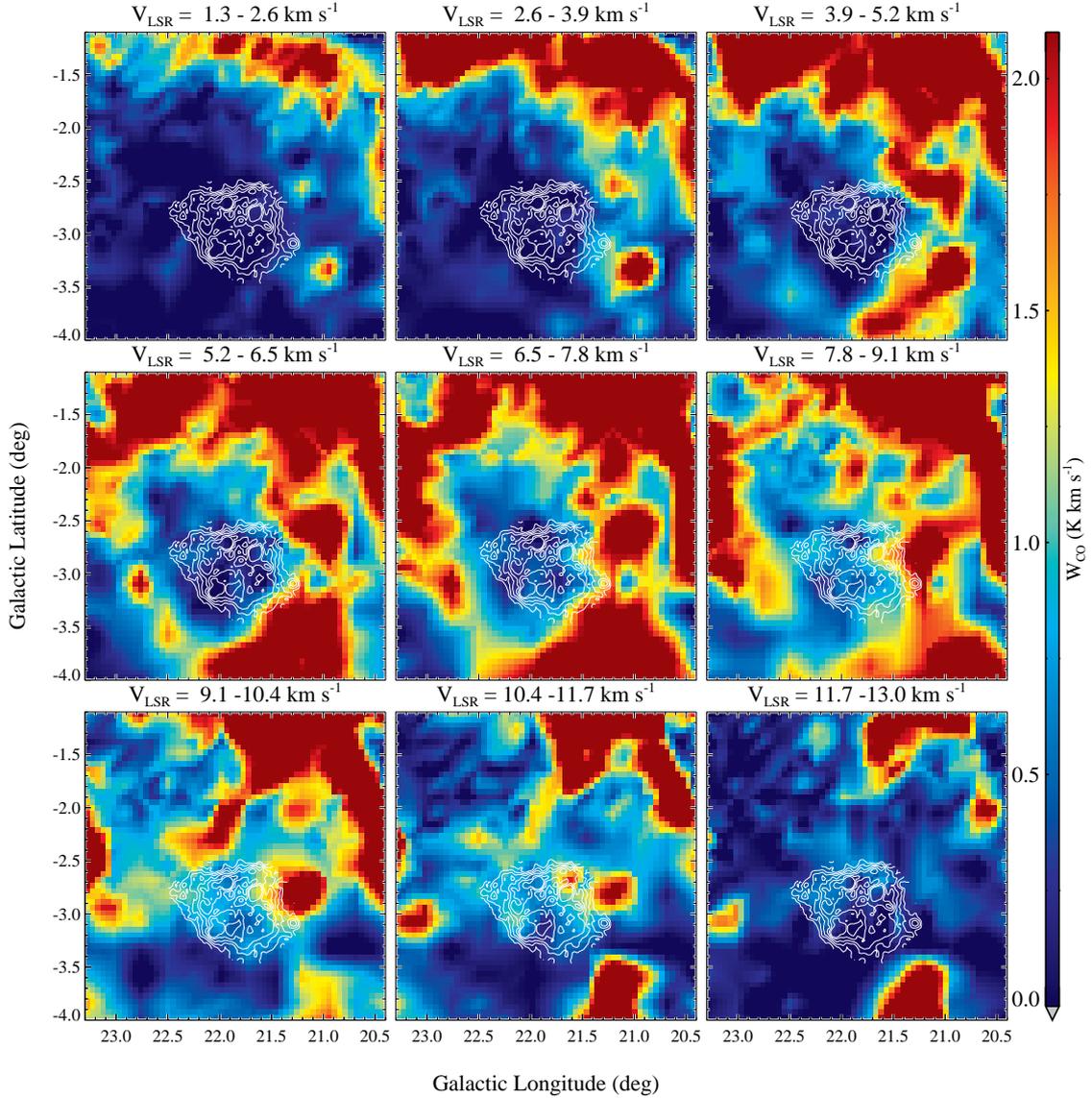}\\
  \vspace{5mm}
  \caption{Intensity maps of $^{12}$CO($J=1-0$) in the area of
    G21.8$-$3.0 showing integrated emission in 1.3~km~s$^{-1}$ wide
    channels for the velocity range of 1.3 $-$ 13.0~km~s$^{-1}$,
    overlaid by contours of total intensity $I$ of Effelsberg
    $\lambda$11\ cm observations.}
\label{channel-maps}
\end{figure*}

No X-ray counterpart was found for G21.8$-$3.0 from the ROSAT X-ray
All-Sky Survey data in both low (0.1 -- 0.4~keV) and high (0.4 --
2.4~keV) bands \citep{Voges99} and the archived data from the Chandra
X-ray observatory. We also failed to find any discrete source that
might be associated with G21.8$-$3.0 in the high-energy Galactic plane
surveys from HESS and HAWC \citep{Jardin19} and the Fermi 4FGL
catalogue \citep{4FGL}.

\subsection{Possibly associated gas structure and the distance}

Among the 294 known Galactic SNRs \citep{Green19}, 70 of them have
physical contact with the surrounding molecular clouds \citep[MCs,
  e.g.][]{cjz+14}. The kinematic distance therefore can be estimated
from the systematic velocity of the associated MCs
\citep[e.g.][]{Foster13}. We explored the possible SNR-MC association
i.e. morphological agreement or corresponding MC features
(e.g. cavities, bubbles) with that of G21.8$-$3.0.

\begin{figure}
  \centering
  \includegraphics[width=0.45\textwidth]{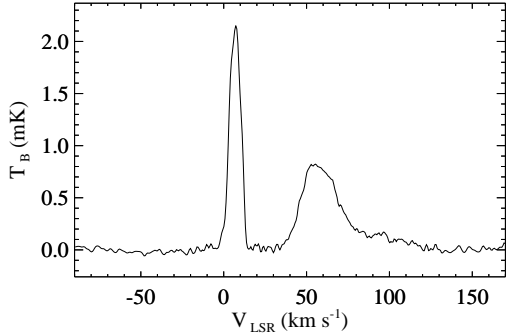}\\
  \caption{Averaged spectrum of $^{12}$CO($J=1-0$) for the area shown
    in Fig.~\ref{channel-maps}.}
\label{co-spec}
\end{figure}

The $^{12}$CO($J=1-0$) survey data from \citet{dht01}, which have an
angular resolution of $\sim$8$\farcm$4 and a velocity resolution of
0.65~km~s$^{-1}$ \citep[see Table~1 of][]{dht01}, are adopted in this
work. As shown in Fig.~\ref{co-spec}, we averaged the CO spectra in
the sky area surrounding G21.8$-$3.0 (see Fig.~\ref{channel-maps}) and
found two major emission features near $V_{\rm LSR} \sim
8$~km~s$^{-1}$ and $\sim$ 55~km~s$^{-1}$, respectively.  By checking
the channel maps around $V_{\rm LSR} \sim 55$~km~s$^{-1}$, the
emission is very extended and originates from the Galactic plane,
i.e. $b > -1\fdg8$, which is 0$\fdg$7 above the radio continuum
emission, hence is not related to G21.8$-$3.0. For the emission
component of $V_{\rm LSR} \sim 8$~km~s$^{-1}$, we show the
$^{12}$CO($J=1-0$) intensity maps integrated in the velocity range of
1.3 $-$ 13.0~km~s$^{-1}$ in steps of 1.3~km~s$^{-1}$ in
Fig.~\ref{channel-maps}. A cavity-like CO feature is seen to encircle
the radio-continuum emission of G21.8$-$3.0 in the velocity range of
$V_{\rm LSR} = 3.9 - 7.8$~km~s$^{-1}$. The southwestern edge of
G21.8$-$3.0 follows the wall of the cavity. The northeastern part of
the cavity extends beyond the continuum emission, and is not
coincident with any discrete sources in the Urumqi $\lambda$6\ cm
image. In the velocity channels of $V_{\rm LSR} < 3.9$~km~s$^{-1}$ or
$>$~7.8~km~s$^{-1}$, the cavity becomes weak and disappears. The
systematic velocity of this cavity-like CO emission peaks at $V_{\rm
  LSR} = 5.9\pm2$~km~s$^{-1}$, which is used to estimate the kinematic
distance of G21.8$-$3.0 by adopting Galaxy rotation models.

Two methods were used. One is the revised kinematic distance
calculator provided by \citet{rmb+14} with the parameters of $R_0 =
8.3$~kpc (distance from the Sun to the Galactic centre), $\Theta_0 =
240$~km~s$^{-1}$ (circular rotation of the local standard of rest),
and the revised Solar motions. The derived nearer kinematic distance
is 0.42$^{+0.61}_{-0.15}$~kpc. The other method is a parallax-based
distance estimator for Galactic sources \citep[][]{rdmb16}: the
estimated distance is 0.25$\pm$0.13~kpc, comparable to that of the
nearest Galactic SNR Vela \citep{Caraveo01}. These two results are
consistent within their errors, locating G21.8$-$3.0 in the Local Arm
\citep[e.g.][]{xhw18}. Based on \citet{Green19}, there are less than
10 SNRs that have a measured distance within 1~kpc. The discovery of
G21.8$-$3.0 would enhance the near-Earth SNR rate and it probably has
made some impact on Earth as discussed in \citet{Firestone14}.

In comparison with the northeastern shell of G21.8$-$3.0, the radio
emission of the southwestern edge, visually contacting the CO clouds,
is relatively weak and does not form a shell, which might imply that
the progenitor star exploded in a pre-existing molecular cavity and
the free-expansion shock just reached or interacted with the wall.

We checked the $\hi$ 21-cm survey data of the Parkes Galactic all-sky
survey \citep[][]{GASS1,GASS2}, which has an effective angular
resolution of $\sim16^\prime$ with a velocity resolution of
1.0~km~s$^{-1}$. The $\hi$ emission in this region is very diffuse and
complex. No related structures can be firmly identified in the
velocity range of $-$50\ km s$^{-1}$ -- 200\ km s$^{-1}$.

\subsection{Size, age and brightness}

Taking the average kinematic distance of 330~pc derived from the CO
data, the diameter of G21.8$-$3.0 is just about 6~pc. Such a small
physical size is an indication of young SNRs. The spectral index of
G21.8$-$3.0, $\alpha = -0.72$, resembles that of some historical SNRs
such as Cas A ($\alpha = -0.77$), SN 1006 ($\alpha = -0.6$) and Tycho
($\alpha = -0.58$) \citep{Green19}. However, unlike the radial
magnetic fields present in these young SNRs \citep{Reynoso13,
  Reich18}, the current images for polarised emission of G21.8$-$3.0
seems to be patchy (see Fig.~\ref{G21.8}). Multi-frequency
polarisation observations with better angular resolution are needed to
clarify the detailed configuration of the magnetic fields in
G21.8$-$3.0.

In the case that G21.8$-$3.0 is not young, the small physical size
requires a low initial explosion energy and/or a high ambient density
into which it expands \citep{Sedov59}. However, as we failed to
identify any related $\hi$ structures, it is difficult to estimate the
initial ambient number density and further the age of G21.8$-$3.0 as
done by \citet{Su17} and \cite{Kothes17}. For the shell-type SNRs,
distance and diameter can also be estimated from the empirical radio
surface brightness -- diameter ($\Sigma$-D) relation. By collecting 65
shell-type SNRs with direct distance estimates as the calibration
sample, \citet{Pavlovic14} found a new steeper slope for the
$\Sigma$-D relation. Based on this relation, we get a physical size of
more than 40~pc for the case of G21.8$-$3.0, so that the distance of
this SNR is larger than 2.3~kpc, much more distant than the value
obtained from the CO cavity. Note that distances estimated from the
$\Sigma$-D relation have an uncertainty as large as 35\% or more due
to several facts, e.g. the intrinsic properties of SNRs, the
environments where SNRs evolve, and the selection effects of the
calibration sample \citep{Green84, Pavlovic13, Kostic16}. We therefore
emphasise that the distance of G21.8$-$3.0 is still an open issue,
which may be settled by new CO/$\hi$ data with higher angular
resolution and sensitivity or future finding of 1720-MHz OH masers due
to SNR-MC interactions, or even using red clump stars or dust
extinction \citep[e.g.][]{Shan18, ycjz19}.

\subsection{Associated pulsar?}

We searched in the ATNF pulsar
catalogue\footnote{http://www.atnf.csiro.au/research/pulsar/psrcat/}
for associated pulsars, that may also hint the distance of
G21.8$-$3.0. PSR J1843$-$1113 ($\ell = 22\fdg06, b = -3\fdg40$) with a
distance of 1.26~kpc is outside G21.8$-$3.0 but not far away. In the
case that it is a run-away pulsar from the centre, the proper motion
should confirm this. We found that the measured velocity vector of PSR
J1843$-$1113 \citep{Desvignes16} did not match the presumed path,
i.e. tracing back to the centre of G21.8$-$3.0, therefore, seems not
to be related. PSR J1840$-$1122 \citep[$\ell = 21\fdg56, b =
  -2\fdg74$,][]{Hobbs04} overlaps G21.8$-$3.0. The distance of the
pulsar is about 8.21~kpc, on the Far 3-kpc Arm according to
\citet{Hou14}. No proper motion information is available yet. If this
SNR-pulsar association could be confirmed, the physical size of
G21.8$-$3.0 would be 140~pc, which is exceptionally large. This is
unlikely and the overlap may be just a coincidence.

No pulsars were identified as stated above implying the progenitor of
G21.8$-$3.0 might not be a massive star. No flat-spectrum component
was seen inside G21.8$-$3.0, indicating the non-existence of a pulsar
wind nebula at the current detection sensitivity.

\section{Summary}

Sensitive large-scale radio continuum surveys are crucial for
discovering Galactic SNRs with low surface brightness. By identifying
radio sources in the Sino-German $\lambda$6\ cm polarisation survey of
the Galactic plane, we serendipitously discovered the extended object
G21.8$-$3.0, which has a circular shape with a diameter of about
1$\degr$. To reveal the radio emission properties, new $\lambda$11\ cm
observations were carried out with the Effelsberg 100-m radio
telescope. Together with the Effelsberg $\lambda$21\ cm data, the
radio continuum spectral index was derived to be $\alpha_{\rm
  6cm-21cm} = -0.72\pm0.16$, consistent with that of synchrotron
emission from SNRs. Polarised emission from G21.8$-$3.0 was detected
both at $\lambda$6\ cm and $\lambda$11\ cm. By comparing polarisation
angles, the $RM$ of G21.8$-$3.0 was estimated to be $\sim$80 -- 90
rad\ m$^{-2}$. An H$\alpha$ filament is seen to run along the northern
edge of G21.8$-$3.0. The weak infrared emission of the source
exemplifies its SNR nature. A cavity-like CO feature was found in the
direction of G21.8$-$3.0 at $V_{\rm LSR}$ = 5.9$\pm$2~km~s$^{-1}$. If
related to G21.8$-$3.0, its distance is about 330~pc, which means it
is a local-arm object, however, the $\Sigma$-D relation for shell-type
SNRs results in a much farther distance.

\section*{Acknowledgements}

We thank the anonymous referee for useful comments that improved this
work, and Dr. James Wicker and Dr. David Champion for their kindness
on language editing. We would like to thank Dr. Alex Kraus and
Dr. Peter M{\"u}ller for support with the observations and data
reduction, and also the electronics group at Effelsberg for connecting
the SPECPOL polarimeter with the $\lambda$11\ cm receiver. This
research is based in part on observations with the Effelsberg 100-m
telescope of the MPIfR, Bonn. The Chinese authors are supported by the
National Natural Science Foundation of China (No. 11988101, U1831103,
11933011, 11833009), the National Key R\&D Program of China
(No. 2017YFA0402701), the Open Project Program of the Key Laboratory
of FAST, NAOC, the Chinese Academy of Sciences, and the Partner group
of the MPIfR at NAOC in the framework of the exchange programme
between MPG and CAS for many bilateral visits. XYG acknowledges
financial support from the CAS-NWO cooperation programme. LGH is
grateful for the support by the Youth Innovation Promotion Association
CAS.




\bibliographystyle{mnras}
\bibliography{bbfile.bib} 








\bsp    
\label{lastpage}
\end{document}